\documentclass[fleqn,usenatbib,onecolumn]{rasti}
\usepackage{newtxtext,newtxmath}
\usepackage[T1]{fontenc}
\DeclareRobustCommand{\VAN}[3]{#2}
\let\VANthebibliography\thebibliography
\def\thebibliography{\DeclareRobustCommand{\VAN}[3]{##3}\VANthebibliography}
\usepackage{threeparttable}
\usepackage{graphicx}

\usepackage{textcomp}
\usepackage{upquote}

\usepackage{graphicx}	

\usepackage{amsmath}
\usepackage{dcolumn}
\usepackage[utf8]{inputenc}
\usepackage{float}
\usepackage{url}

\usepackage{color}

\newcommand{\cm}{cm$^{-1}$}

\title[ExoMol data structures]{Data structures for photoabsorption within the ExoMol project}
\author[Jonathan Tennyson et al.] {Jonathan Tennyson\thanks{E-mail: j.tennyson@ucl.ac.uk (JT)}, Marco Pezzella, Jingxin Zhang and Sergei N. Yurchenko\\
Department of Physics and Astronomy, University College London, Gower Street, WC1E 6BT London, United Kingdom}

\date{Accepted XXX. Received YYY; in original form ZZZ}

\pubyear{2023}

\begin{document}
\label{firstpage}
\pagerange{\pageref{firstpage}--\pageref{lastpage}}

\maketitle
\begin{abstract}
    The ExoMol database currently provides comprehensive line lists for modelling the spectroscopic properties of molecules in hot atmospheres.
    Extending the spectral range of the data provided to ultraviolet (UV) wavelengths brings into play three processes not currently
    accounted for in the ExoMol data structure, namely photodissociation, which is an important chemical process in its own right,
    the opacity contribution due to continuum absorption and predissociation which can lead to significant and observable line broadening effects.
 Data structures are proposed which will allow these processes to be correctly captured and the (strong) temperature-dependent effects
 predicted for UV molecular photoabsorption in general and photodissociation in particular to be represented.
\end{abstract}

\section{Introduction}

The analysis of light as function of wavelengths provides a major window on the Universe. To interpret these signals requires appropriate
laboratory data. The ExoMol project \citep{jt528} was established to provide molecular line list for hot astronomical atmospheres.
Up until recently the ExoMol project, in keeping with similar projects such as HITEMP \citep{jt480,jt763}, TheoReTS \citep{TheoReTS} and NASA Ames \citep{21HuScLe}, has presented results as (large) lists of transitions or spectral lines.
Implicitly this assumes that all lines are discrete transitions between bound states (bound -- bound transitions) even though in
some cases the upper states may actually lie in the continuum so that these transitions, while still appearing as lines,
actually represent part of the bound -- free spectrum. The ExoMol project uses a well-defined format for line lists \citep{jt548}
which has been further developed as part of its data releases \citep{jt631,jt810}.

However, a number of recent developments, discussed below, have extended the scope of the ExoMol project and therefore
the data it provides. This has caused us to consider how to generalise the ExoMol data structure to accommodate both the increased
range of data and also its different uses as bound -- free data are important not only for opacities and spectroscopic
models but also, in that they represent a route to photodissociation, which is an important process for chemical models.
Sharp transitions to states lying above the dissociation limit are already starting to be captured by ExoMol 
as part of standard line lists \citep{jt831,jt858}. However, once occupied these above dissociation states can decay either
by emitting a photon, such as UV flourescence which is an important astrophysical process \citep{Lupu_2011,Gerard_2022}, or they can dissociate. The above dissociation region also contains a continuum component to the photoabsorpion caused, for example, by
excitation to dissociative electronically excited states. Currently neither ExoMol, or indeed none of the databases cited above,
capture this component. At the same time we have started to consider the role of photodissociation \citep{jt840,jt865} which
itself usually comprises sharp lines sitting on top of a continuum.  Photoabsorption into the continuum cannot be represented by the current ExoMol line format.
We note that continuous opacities on a grid of temperatures and wavelengths for molecules were generated by \citet{87KuvaTa} and tabulated experimental vacuum ultra violet (VUV) cross sections of molecules are collected in the MPI-Mainz UV/VIS Spectral Atlas  \citep{13KeMoSa} as a continuous spectrum.

A third issue is the representation of the so-called predissociation which occurs when transitions to excited electronic
states which lie above the dissociation limit spontaneously undergo a further process (usually a curve crossing or tunneling) which leads to dissociation. The resulting lines are observed
to be broadened, often significantly, due to the shorter lifetimes associated with predissociating  states. So far while the ExoMol
project has included states in its line lists which are predissociative, it has ignored the important line broadening effects which result from the reduced lifetime associated
with predissociative states. A recent study
by \citet{jt874} of the spectrum AlH in M-dwarf star Proxima Cen highlights problems with this approach. Pavlenko {\it et al.} 
used the ExoMol AlHambra line list for AlH \citep{jt732} to model this spectrum. For the majority of transitions, which do
not show any effects due to predissociation, this line list worked well but it proved to be less accurate for transitions to
predissociating states. Importantly, only the lines which showed broadening, often by as much as 5 cm$^{-1}$, due to predissociation are not saturated in the stellar spectra meaning that it was only by analysing these broadened lines that Pavlenko {\it et al.}
were able to retrieve abundances of AlH. It would therefore clearly be advantageous to include consideration of predissociation
effects in the ExoMol database.

In this research note we propose a generalisation of the current ExoMol format to allow for the various processes discussed
above. At the same time we draw a clear distinction between the photoabsorption data, needed for spectral and opacity models, and
for the data needed for modelling the chemical consequences of photodissociation.

 \section{The present ExoMol data model}

\begin{figure}
\includegraphics[width=0.5\textwidth]{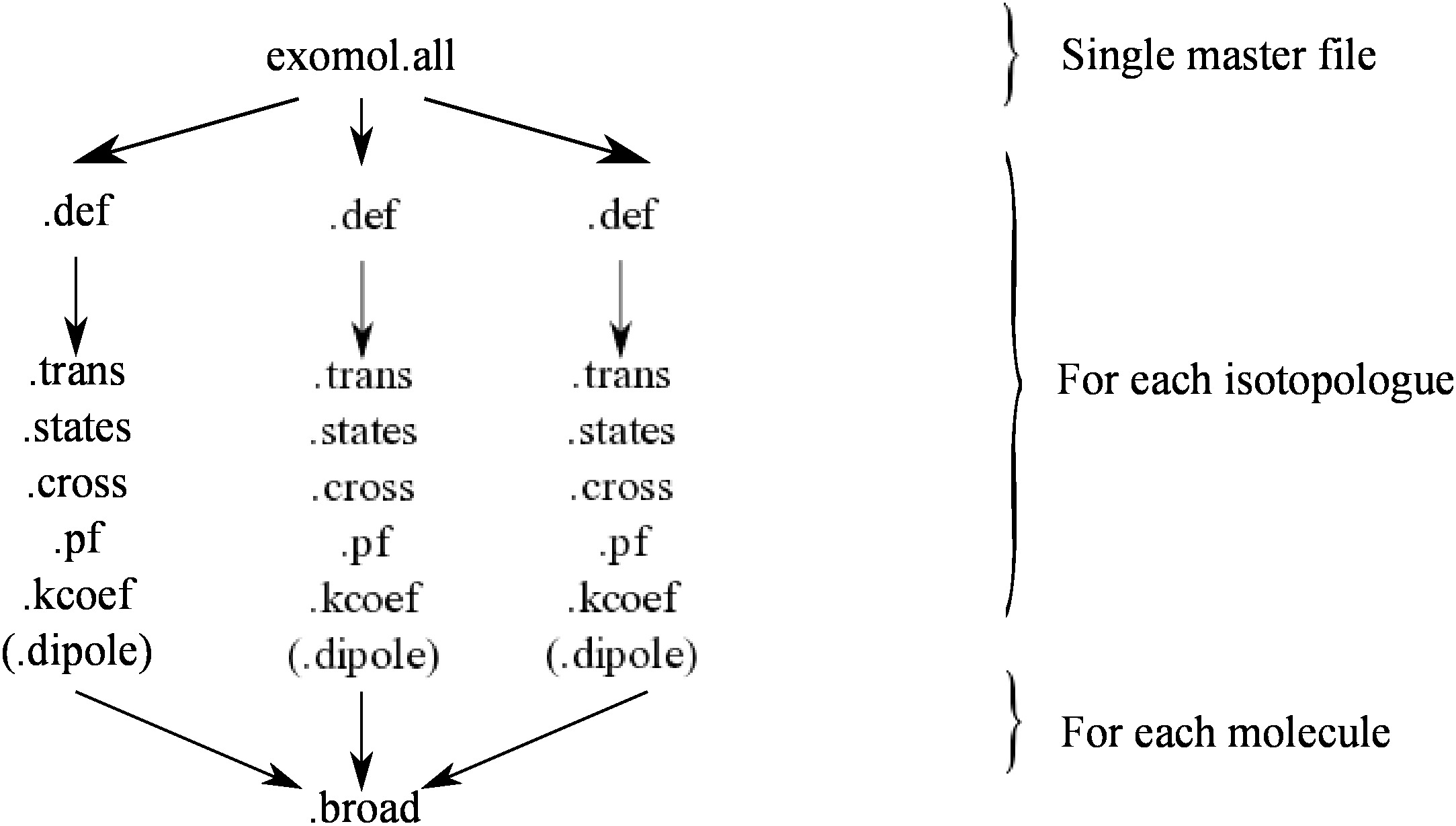}
\caption{Summary of the current ExoMol data structure.}
\label{fig:DS}
\end{figure}

Figure~\ref{fig:DS} gives a simplified ExoMol data structure; a complete specification of the file types is given in Table~\ref{tab:files}.
The master file \texttt{exomol.all} (\url{https://www.exomol.com/exomol.all}) gives an overview of the entire database and points towards the \texttt{.def} files which characterise
the recommended line lists for each isotopologue for which data is available. The \texttt{.def} file contains specification of the dataset in terms
of what is available, for example uncertainties or lifetimes, quantum numbers used in the states file and file sizes. It also gives
a version number in yyyymmdd date format.
 The core of the database are the \texttt{.states} and \texttt{.trans} files
which provide a compact form of the line list data.

\begin{table}
\caption{Specification of the ExoMol file types. (Contents in brackets are
optional.) The final three entries are new file types, see Tables~\ref{tab:cross}, \ref{tab:pdef} and \ref{tab:photo}, respectively, for specifications.}
\label{tab:files}
\tiny
\begin{tabular}{lcll}
\hline\hline
File extension & $N_{\rm files}$&File DSname &  Contents\\
\hline
\texttt{.all} &1& Master& Single file defining contents of the ExoMol
database.\\
\texttt{.def} &$N_{\rm tot}$& Definition& Defines contents of other files for
each isotopologue.\\
\texttt{.states} &$N_{\rm tot}$& States & Energy levels, quantum numbers,
lifetimes, (Land\'e $g$-factors, Uncertainties).\\
\texttt{.trans} &$^a$& Transitions & Einstein $A$ coefficients, (wavenumber).\\
\texttt{.broad} &$N_{\rm mol}$& Broadening & Parameters for pressure-dependent
line profiles.\\
\texttt{.cross}&$^b$& Cross sections& Temperature or temperature and pressure-dependent
cross sections.\\
\texttt{.kcoef}&$^c$& $k$-coefficients& Temperature and pressure-dependent
$k$-coefficients.\\
\texttt{.pf}& $N_{\rm tot}$&Partition function&   Temperature-dependent
partition function,
(cooling function).\\
\texttt{.dipoles}&$N_{\rm tot}$& Dipoles & Transition dipoles including
phases.\\
\texttt{.super} &$^d$& Super-lines & Temperature dependent super-lines (histograms) on a wavenumber grid.\\
\texttt{.nm} &$^e$ & VUV cross sections& Temperature and pressure dependent VUV cross-sections (wavelength, nm).\\
\texttt{.fits}, \texttt{.h5}, \texttt{.kta}  &$^f$ & Opacities & Temperature and pressure dependent  opacities for radiative-transfer applications.\\
\hline
\texttt{.overview}&$N_{\rm mol}$& Overview & Overview of datasets available.\\
\texttt{.readme}&$N_{\rm iso}$& Readme & Specifies data formats.\\
\texttt{.model}&$N_{\rm iso}$& Model & Model specification.\\
\hline
\texttt{.cont}&$N_{\rm iso}/0$&Continuum& Continuum contribution to the photoabsorption.\\
\texttt{.pdef}&$N_{\rm tot}$& Definition& Defines contents of \texttt{.photo} files for
each isotopologue.\\
\texttt{.photo}&$N_{\rm iso}$&Photodissociation& Photodissociation cross sections for each lower state.\\
\hline\hline
\end{tabular}

\noindent
$N_{\rm files}$ total number of possible files;\\
$N_{\rm mol}$ Number of molecules in the database;\\
$N_{\rm tot}$ is the sum of $N_{\rm iso}$ for the $N_{\rm mol}$ molecules in the
database;\\
$N_{\rm iso}$ Number of isotopologues considered for the given molecule.\\
$^a$ There are $N_{\rm tot}$ sets of \texttt{.trans} files but for molecules
with large
numbers of transitions the \texttt{.trans} files are subdivided into wavenumber
regions.\\
$^b$ There are $N_{\rm cross}$ sets of \texttt{.cross} files for
isotopologue.\\
$^c$ There are $N_{\rm kcoef}$ sets of \texttt{.kcoef} files for  each
isotopologue.\\
$^d$ There are $N_T$ sets of $T$-dependent super-lines.  \\
$^e$ There are $N_{VUV}$ sets of VUV cross sections. \\
$^f$ Set of opacity files in the format native to specific radiative-transfer programs. 
\end{table}

\begin{table}
\caption{Specification of the mandatory part of the states file with 
extra data options unc, $\tau$ and $g$.}
\label{Tab:states}
\begin{tabular}{llll} \hline
Field & Fortran Format & C Format & Description\\ \hline
$i$ & \texttt{I12} & \texttt{\%12d} & State ID\\
$E$ & \texttt{F12.6} & \texttt{\%12.6f} & State energy in $\mathrm{cm^{-1}}$\\
$g_\mathrm{tot}$ & \texttt{I6} & \texttt{\%6d} & State degeneracy\\
$J$ & \texttt{I7/F7.1} & \texttt{\%7d/\%7.1f} & $J$-quantum number (integer/half-integer)\\
(unc) & \texttt{F12.6} & \texttt{\%12.6f} & Uncertainty in state energy in $\mathrm{cm^{-1}}$ (optional) \\
($\tau$) & \texttt{ES12.4} & \texttt{\%12.4e} & Lifetime in s (optional)\\
($g$) & \texttt{F10.6} & \texttt{\%10.6f} & Land\'e $g$-factor  (optional)\\
\hline
\end{tabular}

\noindent
 {\flushleft
ID: state identifier: a non-negative integer index, starting at 1\\
$J$:  total angular momentum quantum,  excluding nuclear spin \\

Fortran format, $J$ integer: \texttt{(I12,1x,F12.6,1x,I6,I7,1x,ES12.4,1x,F10.6)}\\
or $J$ half-integer:  \texttt{(I12,1x,F12.6,1x,I6,F7.1,1x,ES12.4,1x,F10.6)}\\
}

\end{table}

Table~\ref{Tab:states} gives the specifications for the mandatory part of the \texttt{.states} file.  These specifications include the optional components:
 uncertainties in the term values, the state lifetime and Land\'e $g$-factor. The specification of term value uncertainties was introduced
as part of the last data release \citep{jt810} and the aim is to make their inclusion compulsory once the available datasets have all had
uncertainties added. The lifetimes column has thus far contained radiative lifetimes computed using the Einstein A coefficients available
in the transitions file \citep{jt624}; so far lifetime effects due to other processes such as predissociation have not been considered. As discussed
below we propose changing this.

After the mandatory fields, the states files contains data on the quantum numbers and other meta-data used to specify each state. The number and format of these
quantum numbers is specified in by the \texttt{.def} file associated with that dataset. Other metadata associated with level can also be
included in this section. Table~\ref{tab:AlOstates} shows the start of the states file for the recently update AlHambra line list
for $^{27}$Al$^{16}$O \citep{jt835}. Note that we have taken the opportunity to update
our quantum number specifications to conform with the  \texttt{PyValem}  python package for parsing, validating, manipulating quantum states labels of atoms, ions and small molecules \citep{pyvalem,pyvalem2}. In general, this change only affects
electronic state designations for which \texttt{X2SIGMA+}, \texttt{A2PI}
and soforth are updated to \texttt{X(2SIGMA+)}, \texttt{A(2PI)}, etc.. This update adds two
characters to the electronic state field but otherwise should be transparent to users; however,
it means that all state labels can now be parsed using  \texttt{PyValem} which is important
for some uses of the database \citep{jtLiDB}.

Table~\ref{tab:trans} gives the specification of the simpler but generally much larger \texttt{.trans} file.

\begin{table*}
    \centering
    \caption{An excerpt from the recently updated states file for $^{27}$Al$^{16}$O, see \citet{jt835}.}
    \label{tab:AlOstates}
    {\tt
    \begin{tabular}{rrcccccclccccc}
        \hline
        $i$ & $\tilde{E}$ ({\rm cm}$^{-1}$)& $g$ & $J$ & \multicolumn{1}{c}{\rm unc (cm$^{-1}$) }& $\tau$ {\rm (s)} & $+/-$ & {\rm e/f} & {\rm State} & $\varv$ & |$\upLambda$| & |$\upSigma$| & |$\upOmega$| & {\rm Source Label} \\
        \hline
        1 & 0.000000 & 12 & 0.5 & 0.000001 & inf & + & e & X(2SIGMA+) & 0 & 0 & 0.5 & 0.5 & EH \\
        2 & 965.416878 & 12 & 0.5 & 0.001651 & 3.6193e+01 & + & e & X(2SIGMA+) & 1 & 0 & 0.5 & 0.5 & PS \\
        3 & 1916.827286 & 12 & 0.5 & 0.010519 & 9.1761e+00 & + & e & X(2SIGMA+) & 2 & 0 & 0.5 & 0.5 & PS \\
        4 & 2854.162814 & 12 & 0.5 & 0.022148 & 4.2147e+00 & + & e & X(2SIGMA+) & 3 & 0 & 0.5 & 0.5 & PS \\
        5 & 3777.464572 & 12 & 0.5 & 0.019275 & 2.3874e+00 & + & e & X(2SIGMA+) & 4 & 0 & 0.5 & 0.5 & PS \\
        6 & 4686.658929 & 12 & 0.5 & 0.016235 & 1.2041e+00 & + & e & X(2SIGMA+) & 5 & 0 & 0.5 & 0.5 & PS \\
        7 & 5346.089546 & 12 & 0.5 & 0.009700 & 2.0501e-04 & + & e & A(2PI) & 0 & 1 & 0.5 & 0.5 & Ma \\
        8 & 5581.908884 & 12 & 0.5 & 0.014747 & 4.3685e-02 & + & e & X(2SIGMA+) & 6 & 0 & 0.5 & 0.5 & PS \\
        \hline
    \end{tabular}
    }
    \begin{tablenotes}
        \item $i$: State counting number;
        \item $\tilde{E}$: Term value (in cm$^{-1}$);
        \item $g_{\rm tot}$: Total state degeneracy;
        \item $J$: Total angular momentum quantum number;
        \item unc: Estimated uncertainty of energy level (in cm$^{-1}$);
        \item $\tau$: Lifetime (in s$^{-1}$);
        \item $+/-$: Total parity;
        \item e/f: Rotationless parity;
        \item State: Electronic term value;
        \item $\varv$: Vibrational quantum number;
        \item |$\upLambda$|: Absolute value of the projection of electronic angular momentum; 
        \item |$\upSigma$|: Absolute value of the projection of the electronic spin;
        \item |$\upOmega$|: Absolute value of the projection of the total angular momentum;
        \item Source Label: Method used to generate term value \citep{jtquad}.
    \end{tablenotes}
\end{table*}

\begin{table}
\caption{Specification of the transitions file.}
\label{tab:trans}
\begin{tabular}{llll} \hline
Field & Fortran Format & C Format & Description\\ \hline
$i$ & \texttt{I12} & \texttt{\%12d} & Upper state ID\\
$f$ & \texttt{I12} & \texttt{\%12d} & Lower state ID\\
$A$ & \texttt{ES10.4} & \texttt{\%10.4e} & Einstein $A$ coefficient in
$\mathrm{s^{-1}}$ \\
$\tilde{\nu}_{fi}$&\texttt{E15.6} & \texttt{\%15.6e} & Transition wavenumber in
cm$^{-1}$ (optional). \\
 \hline
\end{tabular}

\noindent
Fortran format: \texttt{(I12,1x,I12,1x,ES10.4,1x,ES15.6)}\\
\end{table}

\section{Proposed updated ExoMol data model}

\begin{figure}
    \centering
    \includegraphics[width=0.7\textwidth]{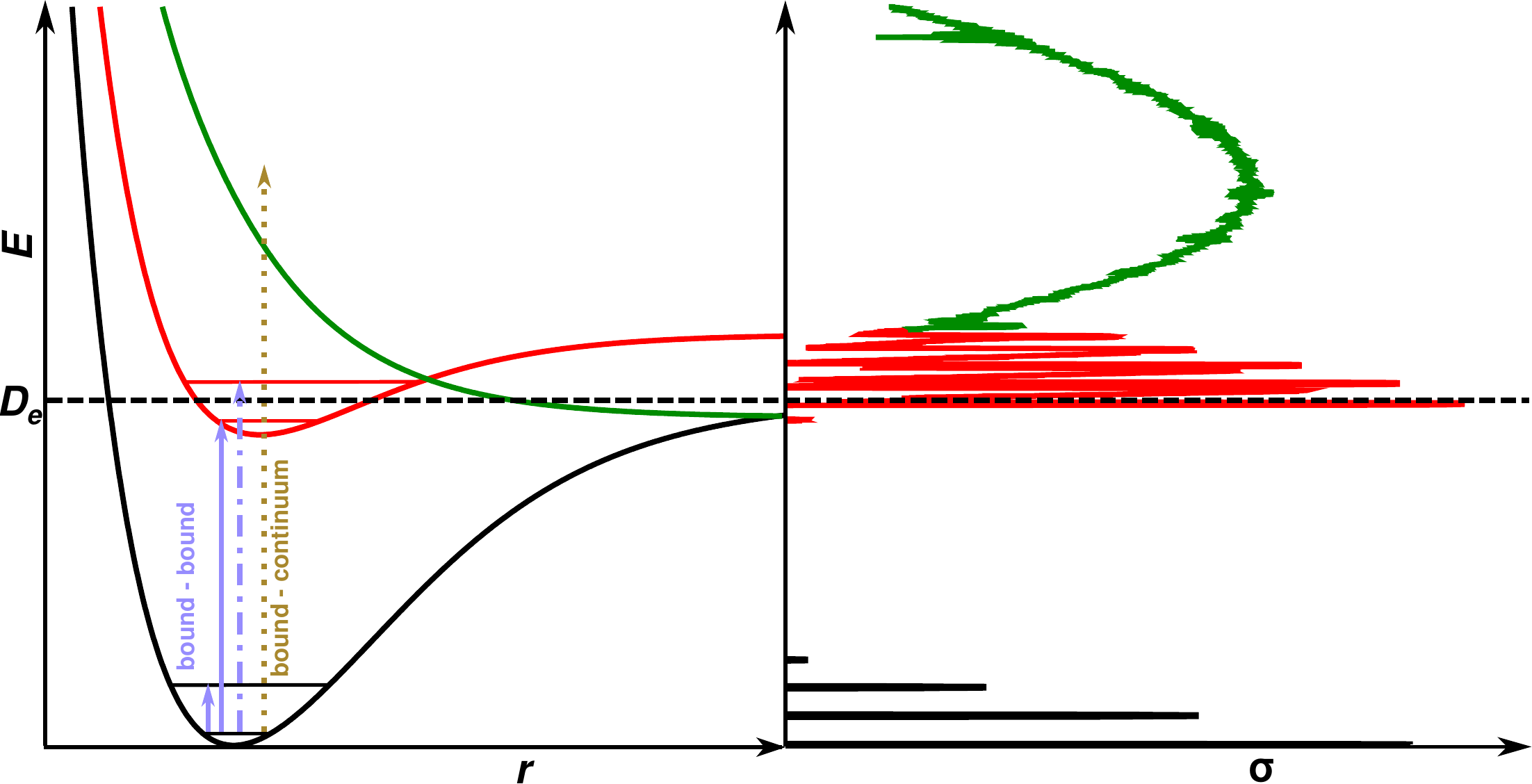}
    \caption{\label{f:PECs:HF} Schematic representation of photoabsorption for a diatomic molecule showing line and  continuum spectra. The transitions within the ground state electronic state (in black) lead to the rovibrational line spectrum (also in black) represented by transitions to  excited electronic states (in red); transitions to the repulsive electronic  state produce continuum spectra (in green). The line arrows denote sharp, line transitions, the dot interrupted arrow goes to states  above the  dissociation limit ($D_{\rm e}$) which can exhibit predissociative effects. The golden dotted arrow shows photoabsorption which we represent using bound-continuum cross sections. Photoabsorption above $D_{\rm e}$ can contribute to photodissociation.} 
\end{figure}

There are three new aspects that need to be included in an updated ExoMol data structure: predissociation, the continuum contribution to the opacity and
photodissociation. Figure~\ref{f:PECs:HF} illustrates the various photoabsorption processes for the case of a diatomic molecule.

\subsection{Predissociation}
\begin{figure}
    \centering
    \includegraphics[height=6cm]{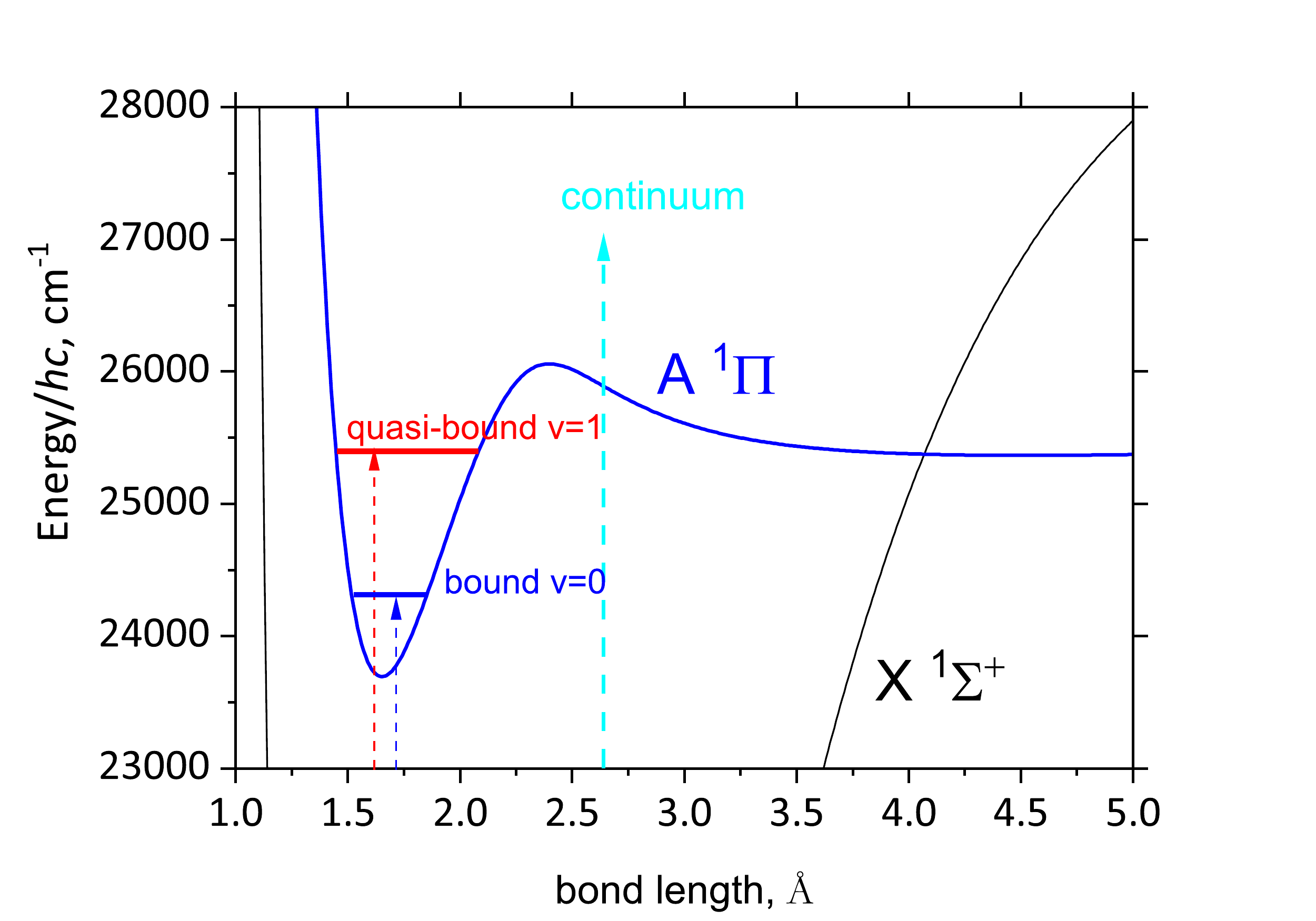}
\includegraphics[height=6cm]{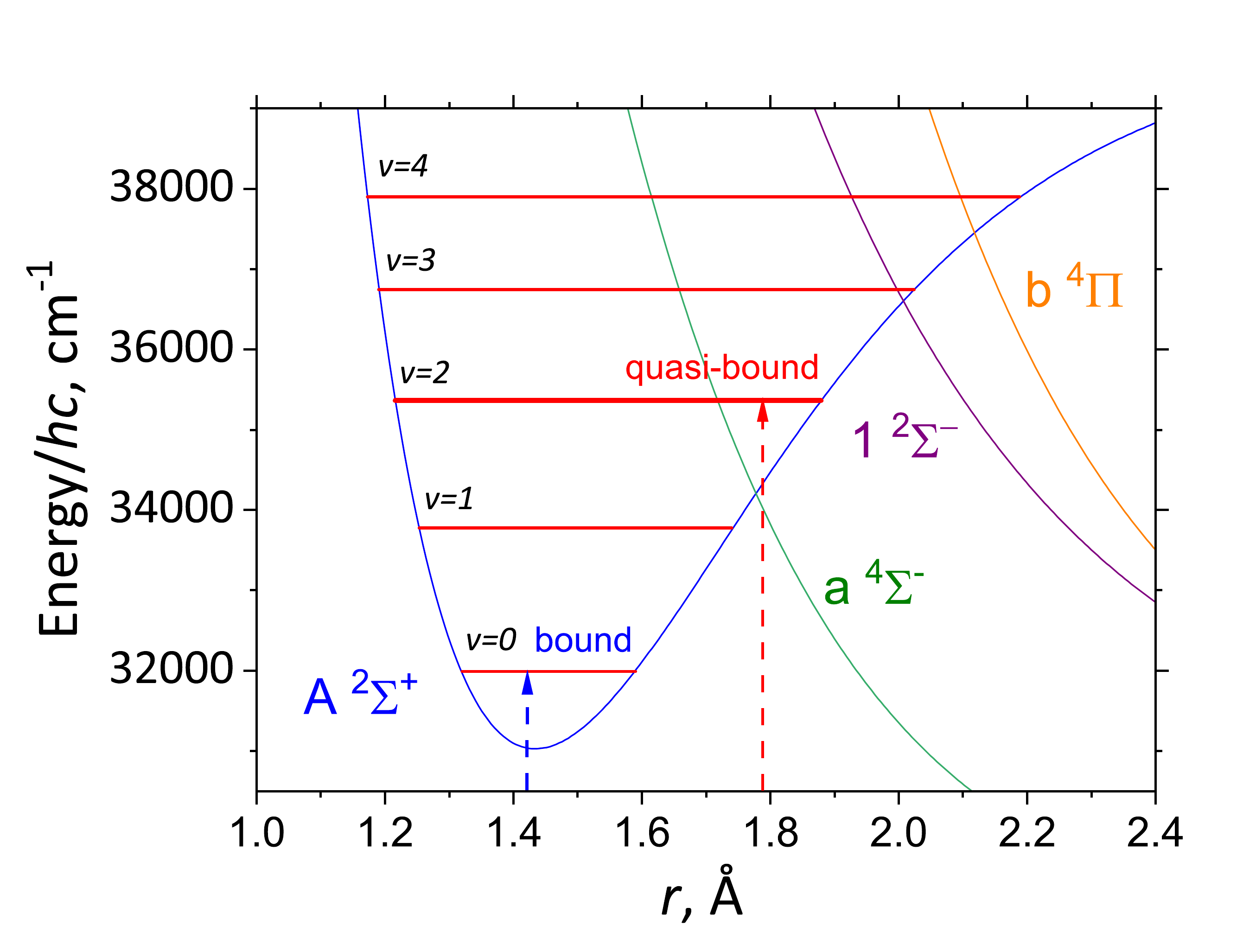}
    \caption{\label{f:PECs:AlH} Left: Potential energy curves for AlH due to \citet{jt732} showing the predissociation region. The $v=0$ vibrational state of the A~$^1\Pi$  electronic
    state lies below the AlH dissociation limit: transitions to this state are sharp as they do not show effects due to predissociation. Conversely, states in the $v=1$ can predissociate by tunneling through the barrier to dissociation and show pronounced effects due to lifetime broadening. Right: The potential energy curves (bound and dissociative) and predissociated states of SH.}
\end{figure}

Figure~\ref{f:PECs:AlH} illustrates the main mechanism leading to predissociation; for AlH it is caused by tunneling through a small barrier to a dissociation, while for SH it is caused by couplings to dissociative states crossing the state to which the transition goes to. The effect of predissociation can be included by a minor adjustment to the current ExoMol data structure. Predissociation manifests as a shortened lifetime which leads to enhanced natural line-broadening of any transition to (or from) the state concerned. 
We therefore propose to generalise the definition of the lifetime, $\tau$, given in the \texttt{.states} file.
For most line lists, where predissociation is not important, $\tau$ is defined as the natural lifetime due to 
radiative decay. 
In cases where predissociation is considered,  $\tau$ will represent the natural life due to both 
radiative decay and predissociation.  For example, the radiative  lifetime of $v=1$ A~$^2\Sigma^+$ state of SH  was calculated by \citet{jt776} as 5.13 ns, while the predissociative lifetimes was measured as 5.45(24)~ps ($N=0$) \citep{97WhOrAs.SH}. 
A similar example of the predissociation effects in the spectra of AlH is the predissociative lifetime of the $J=23$, $v=1$, A $^1\Pi$ state of $^{27}$AlH, which  was measured by \citet{79BaNexx.AlH} as 4.5~ns, while the radiation lifetime is calculated to be 101.7~ns \citep{jt732}. The reduced lifetime affects the line broadening of the corresponding transitions and therefore is an important factor in retrievals of AlH abundance from stellar spectra, as was recently demonstrated by \citet{jt874}. 

Cases where predissociation effects are included in this lifetime will be marked
by a new flag, \texttt{prediss}, included in the \texttt{.def} file; \texttt{prediss} will default to 0 (none) and be set to 1 when the effects are present.

In principle, the natural lifetime provides a contribution to the line profile which sits alongside the temperature-dependent
effects of Doppler broadening and pressure-dependent collisional broadening. In practice, the natural lifetime usually makes
a minimal contribution to the line profile and is thus neglected. For predissociating states this is no longer the case.
However, including the effect of lifetime broadening within a standard Voigt profile is straightforward. Lifetime broadening leads to a Lorentzian line shape like pressure broadening where $\gamma_\tau$, the half-width in cm$^{-1}$ due lifetime broadening,
is given by 
\begin{equation}
\gamma_\tau = \frac{\hbar}{2\tau h c}.   
\end{equation} 
This half-width can simply be added to the pressure-broadening half-width, $\gamma_p$, to give the total Lorentzian component of the
line profile. Given that a Voigt profile is already being used, this has little computational impact on a calculation which 
suggests that routine use of  
$\gamma_p+\gamma_{\tau}$ for half-width would avoid the need to worry about whether predissociation needs to be
considered or not. To this end, for states with non-negligible predissociative lifetimes, the radiative values of $\tau_{\rm rad}$ in the ExoMol States files will be replaced by $\tau_{\rm prediss}$. In the  example of the $J=23$, $v=1$, A $^1\Pi$ state of $^{27}$AlH, the ExoMol value $\tau_{\rm rad}=1.0169\times10^{-7}$~s can be replaced by the experimental value $\tau_{\rm prediss}=4.5\times10^{-9}$~s \citep{79BaNexx.AlH}, otherwise calculated values will be used.

\subsection{Continuum absorption}

Both continuum absorption and photodissociation need to be represented as cross sections rather than lines as they are continuum processes. However, in our proposed
data model continuum absorption will be represented as a function of wavenumber (cm$^{-1}$) to retain consistency with the line spectra, while photodissication
cross sections will be stored as function of wavelength (nm).

When considering photoabsorption by molecules to states lying above the dissociation limit, the spectrum can be thought of broadly dividing into two classes: line spectra comprising what looks like sharp bound-bound transitions, and absorption directly into the continuum. Predissociation spectra form an intermediate between these two cases and, as discussed above, will be treated as line spectra. The line spectra can be represented using the form of the standard line lists (line positions, Einstein A coefficients, lower/upper state energies and other state descriptions), as captured by the \texttt{.states} and \texttt{.trans} files. However, the bound-continuum photoabsorption is best represented as temperature-dependent photo-absorption cross-sections. 
These continuum photoabsorption cross sections, whose data specification is given below, will be stored as part of the standard line list data base as separate files for each isotopologue. The temperature-dependent photoabsorption spectrum is then obtained by adding the appropriate line and continuum
cross-sections using software such as \texttt{ExoCross}  \citep{jt708} or \texttt{PyExoCross} \citep{jtPyExoCross}.

Continuum molecular absorption due to collision induced absorption (CIA) \citep{12RiGoRo.CIA,19KaGoVa.CIA} is already routinely considered as part of astronomical models; it has also been recently recommended that absorption due to the so-called \lq water continuum\rq\ be considered in model atmospheres for water-rich exoplanets \citep{jt850}.

The data model we propose for including continuum absorption due to simply photoabsorption into continuum states is given in Table~\ref{tab:cross}. Again these cross-sections will be temperature dependent but,
unlike CIA and the water continuum, it is probably safe to assume that this continuum is not strongly pressure dependent. The presence of a continuum absorption
contribution to the photoabsorption will be indicated 
by a new flag, \texttt{continuum}, included in the \texttt{.def} file; \texttt{continuum} will default to 0 (none) and be set to 1 when the effects are present.

We note that our proposal involves providing photodissociation data on a wavelength grid (in nm) while continuum absorption cross sections will be provided on a wavenumber grid (in cm$^{-1}$). This latter choice ties in closely with line lists which already prove transitions wavenumber (in  cm$^{-1}$). The data structure of continuum absorption cross sections is presented in  Table~\ref{tab:cross}. The file names have the
following structure:
\texttt{\textquotesingle{}<ISO-SLUG>\_\_<DSNAME>\_\_<RANGE>\_\_T<TEMP>K\_\_P<PRESSURE>bar\_\_<STEP>.cont\textquotesingle{}},\\
where \texttt{ISO-SLUG} is the iso-slug molecule name \citep{jt810}, \texttt{DSNAME} is the name of the line list, \texttt{RANGE} is the wavenumber range in \cm, \texttt{TEMP} is the temperature in K, \texttt{PRESSURE} is the pressure in bar, \texttt{STEP} is the wavenumber step in \cm.

\begin{table}
\caption{Specification of the \texttt{.cont} continuum
photoabsorption cross section file format.}\label{tab:cross}
\begin{tabular}{llll}
\hline\hline
Field & Fortran Format & C Format & Description\\
\hline
$\tilde{\nu}_i$ & \texttt{F12.6} & \texttt{\%12.6f} & Central bin wavenumber, cm$^{-1}$\\
$\sigma_i$ & \texttt{ES14.8} & \texttt{\%14.8e} & Absorption cross section,
$\mathrm{cm^2\,molec^{-1}}$\\
\hline\hline
\end{tabular}

\noindent
Fortran format: \texttt{(F12.6,1x,ES14.8)}\\
\end{table}

\subsection{Photodissociation}

Photodissociation cross sections are separated from the line list and form a distinct section in the ExoMol database. At present this section
contains calculated cross sections for HCl and HF \citep{jt865} and measured cross sections for CO, H$_2$O, CO$_2$, SO$_2$, NH$_3$, H$_2$CO and C$_2$H$_4$ due
to \citet{jtVUV} and CO$_2$ due to \citet{18VeBeFa.CO2}. The immediate plan is also to add temperature-dependent cross sections due to
Qin and co-workers who have performed photodissociation calculations on  MgO \citep{21BaQiLi.MgO}, AlH \citep{21QiBaLi.AlH}, AlCl \citep{21QiBaLi.AlCl},
AlF \citep{22QiBaaLi.AlF} and O$_2$ \citep{23HuQiNa.O2}, as well as HF and HCl \citep{22QiNaLi.HCl}. In due course a structure of  photodissociation \texttt{.pdef} files will be implemented to aid the navigation of this section of the database.

As the photodissociation cross sections form a distinct part of the ExoMol database, we have added a new photodissociation definition file (\texttt{.pdef}) file to the
data structure; the proposed format of this file is given in Table~\ref{tab:pdef}. This gives the necessary metadata to access and interpret the recommended photodissociation cross sections. The information section mirror those given in the \texttt{.def} for the same system.
For completeness we have added two more flags to the master file, \texttt{line} and \texttt{photo}, which define whether the line spectra and photodissociation cross sections
are present (=1) or not (=0). The default values are \texttt{line}=1 and \texttt{photo}=0 which aligns with structure of previous master files which assumed all data was in the form
of a line list.

A file structure for photodissociation was already proposed by \citet{jt865}; however, this is updated and extended here to align with the one proposed for VUV spectra
in \citet{jt810}; Table~\ref{tab:photo} gives the formal specification of the file
structure. As a file naming convention we adopt the following:\\
\texttt{\textquotesingle{}<ISO-SLUG>\_\_<DSNAME>\_\_<RANGE>\_\_T<TEMP>K\_\_P<PRESSURE>bar\_\_<STEP>.photo\textquotesingle{}}, \\
where \texttt{ISO-SLUG} is the iso-slug molecule name, \texttt{DSNAME} is the name of the line list, \texttt{RANGE} is the wavelength range in nm, \texttt{TEMP} is the temperature in K, \texttt{PRESSURE} is the pressure in bar, \texttt{STEP} is the wavelength step in nm. For example, the states file of the photodissociation cross sections for HF the filename: \url{1H-19F__PTY__90.0-400.1__T200K__P0bar__0.1.nm}, see Table~\ref{tab:nm}.

\citet{jt865} found that their cross sections depended strongly on the temperature of the molecule and proposed presenting these data for 34 temperatures between $T = 0$ and $T = 10 000$~K. This data model implicitly assumes that the molecule is in local thermodynamic equilibrium (LTE). We discuss issues with treating non-LTE effects and other issues
with this data model in the next section. These data are all implicitly at zero pressure as pressure broadening effects are neglected. Data from other sources
will have different temperature and pressure grids.


\begin{table}
\caption{Photodissociation definition  (\texttt{.pdef}) file format; each entry starts on a new line.} 
  \label{tab:pdef} 
\begin{tabular}{lrll}
\hline 
\multicolumn{1}{c}{Field} & Fortran Format & C Format &   Description \\
\hline
\multicolumn{4}{l}{\textbf{Header Information}} \\
\hline
\texttt{ID} & A11  & \%11s & Always the ASCII string ``EXOMOL.pdef'' \\
\texttt{IsoFormula} & A27 & \%27s & Isotopologue chemical formula\\
\texttt{Iso-slug} &  A160 & \%160s & Isotopologue identifier, see text for
details\\
\texttt{DSName} &  A10 & \%10s & Isotopologue dataset name\\
$V$ & I8  & \%8d & Version number with format YYYYMMDD\\
\texttt{MolKey} &  A27 & \%27s & Standard inchi key of the molecule \\
$N_{\rm atom}$ & I4  & \%4d & Number of atoms\\
\hline
\multicolumn{4}{l}{\textbf{Atom definition} (The following 2 lines occur $N_{\rm atom}$ times)} \\
\hline
$I_{\rm atom}$ & I3   & \%3d & Isotope number \\
Atom & A3 & \%3s & Element symbol\\
\hline
\multicolumn{4}{l}{\textbf{Isotopologue Information}} \\
\hline
$m_{Da} \quad m_{kg}$ & {\small F12.6,1X,ES14.8}   & {\small \%12.6f \%14.8e} & Isotopologue mass
in Da and kg \\
$I_{\rm sym}$ & A6 & \%6s & {\small Molecular symmetry Group (if $N_{\rm atom}=2$ then C or D)}\\
$N_{\rm irrep}$ & I4 & \%4d & Number of irreducible representations\\
\hline
\multicolumn{4}{l}{\textbf{ExoMol Information}} \\
\hline
\texttt{FileType} &  A27 & \%27s & Data source: ExoMol, Expt., etc. \\
$T_{\rm max}$ & F8.2 & \%8.2f & Maximum temperature for cross sections \\
$N_{\rm temp}$ & I3 &   No. of temperatures available\\
$N_{\rm pres}$ & I3 &   No. of pressures available\\
\hline
\hline
\end{tabular}  
\end{table}

\begin{table}
\caption{Specification of the \texttt{.photo} photodissociation cross  section file
format.}\label{tab:photo}
\begin{tabular}{llll}
\hline\hline
Field & Fortran Format & C Format & Description\\
\hline
$\tilde{\lambda}_i$ & \texttt{F10.3} & \texttt{\%10.3f} & Central bin wavelength, nm\\
$\sigma_i$ & \texttt{ES13.6} & \texttt{\%13.6e} &  Photodissociation cross section, $\mathrm{cm^2\,molec^{-1}}$\\
\hline\hline
\end{tabular}

\noindent
Fortran format: \texttt{(F10.3,1x,ES13.6)}\\
\end{table}

\begin{table}
\caption{An excerpt from the \texttt{.nm} photodissociation file for $^1$H$^{19}$F, see \citet{jt865}.}
\label{tab:nm}
\setlength{\tabcolsep}{8mm}{
\begin{tabular}{cc} \hline
$\tilde{\lambda}_i$ & $\sigma_i$ \\ \hline
90.00 & 4.48427452E-19 \\ 
90.10 & 4.50720456E-19 \\ 
90.20 & 4.52964494E-19 \\ 
90.30 & 4.55158741E-19 \\ 
90.40 & 4.57293432E-19 \\ 
90.50 & 4.59377114E-19 \\ 
90.60 & 4.62624329E-19 \\ 
90.70 & 1.68471832E-18 \\
\hline
\end{tabular}
}
\noindent
 {\flushleft
$\tilde{\lambda}_i$: Central bin wavelength, nm; \\
$\sigma_i$:  Photodissociation cross section, $\mathrm{cm^2\,molec^{-1}}$. \\
}

\end{table}

\begin{figure}
    \centering
    \includegraphics[width=0.7\textwidth]{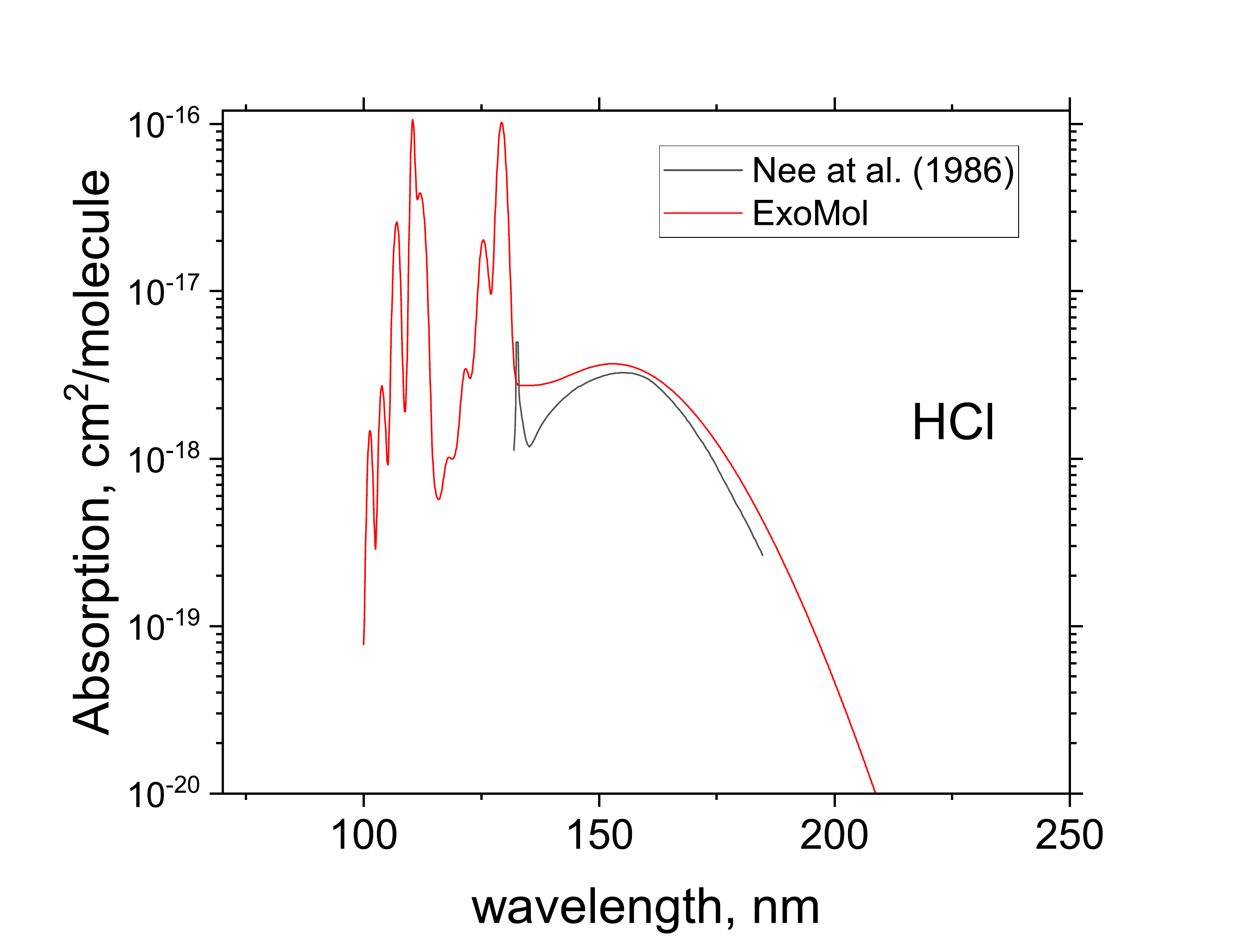}
    \caption{\label{f:HCl:photo} Photodissociation spectra of HCl from ExoMol \citep{jt865}  (theoretical, using natural abundance) and by \citet{86NeSuMa.HCl} (298 K) as provided by the MPI-Mainz UV/VIS Atlas.} 
\end{figure}

Experimental cross sections of molecules covering the VUV region has been curated by the MPI-Mainz UV/VIS Spectral Atlas \citep{13KeMoSa} using a similar, wavelength (in nm) format. In Fig.~\ref{f:HCl:photo}, we illustrate photodissocitasion cross sections of HCl from ExoMol \citep{jt865} and by \citet{86NeSuMa.HCl} at room temperature as provided by the MPI-Mainz UV/VIS Atlas.

\section{Omissions from the updated data model}

 The assumption of LTE for a molecule undergoing photodissociation may not be valid in all cases. Our method of calculating these cross sections does indeed involve computing the initial/final-state dependent data which would be required for a non-LTE representation of photodissociation. However, given that even for diatomic molecules a large number of initial states have to be considered, even considering the vibronic states only, the volume of these data is large. As yet no-one has asked us for non-LTE photodissociation cross sections so at present we do not propose including them in the standard data distribution; if they are required they can be obtained from the present authors. 
 Examples of the state-dependent non-LTE  cross sections include the continuous opacities of CH and OH provided by \citet{87KuvaTa} as well as the CH data produced by \citet{22PoHoBe} and used in their non-LTE analysis of CH in metal-poor stellar atmospheres. 

 Another issue with our data model for photodissociation is that at present it provides no information of dissociation products. In comparison with
 the Leiden database \citep{17HeBova}, which provides low-temperature photodissociation cross sections for molecules of astrophysical interest, gives the dissociation products
 which are given at the threshold to photodissociation but does not provide information on other possible photodissociation products as they may arise at shorter wavelengths. 
 Although our methodology is capable of providing the partial 
 cross sections (or branching ratios) associated with dissociation to different products, so far our models have not been constructed to produce these data. While the initial step in photodissociation generally involves dipole
 allowed transition to an electronically excited state, the subsequent dissociation step may involve crossings to states which cannot be reached by dipole transitions from the
 ground state such as ones with different spin multiplicity. Allowing for these extra states represents a significant complication in the calculation and again, so far, no one
 has asked for these data. However, should these partial cross sections be needed it would be relatively simple to extend our proposed data model to accommodate them; we note that the International Atomic
 Energy Agency's \texttt{CollisionDB} database \citep{CollisionDB,CollisionDB2} already used \texttt{PyValem} to
 address this issue in collision cross sections.

 \section{Conclusion}
The present research note sets out how we propose to extend the current ExoMol data model to accommodate photoabsorption processes which occur at shorter wavelengths where
the possibility of either direct or indirect absorption into the continuum can occur. Broadly these processes are classified as predissociation, 
continuum absorption and photodissociation contribution to the opacity. While both predissociation and
continuum absorption can be accommodated by generalising our current line; a new branch starting form a photodissociation definition file has been added for the photodissociation cross
sections.

Given that these various processes considered are not mutually exclusive as, for example, photodissociation also provides a contribution to the opacity, some care
is required in defining data structures to facilitate the use of our extended data. We believe our proposals satisfy this requirement but that further expansion will be required
to allow for pressure-dependent continuum absorption, photodissociation of molecules not in LTE, or to account for the possibility that photodissociation might result in 
creation of a variety of different photodissociation products.

\section*{Acknowledgements}

We thank Ahmed Al-Refaie, Richard Freedman, Christian Hill, Roxana Lupu, Zhi Qin, Olivia Venot and Ingo Waldmann for helpful discussions on
the topic of this work. This work was supported by the European Research Council under Advanced Investigator Project 883830.

\section*{Data Availability}
The data discussed in this article is available from the ExoMol database which can be accessed at \url{www.exomol.com}.

\bibliographystyle{rasti}

\end{document}